\documentstyle[aps,prl,multicol,psfig]{revtex}
\begin{document}
\title{Anomalous optical absorption in overdoped cuprates
near the charge-ordering instability}
\author{S. Caprara$^1$, C. Di Castro$^1$, S. Fratini$^{1,2}$, and 
M. Grilli$^1$}
\address{$^1$Istituto Nazionale per la Fisica della Materia - Unit\`a di 
Roma 1, and Dipartimento di Fisica - Universit\`a di Roma ``La Sapienza'',
Piazzale Aldo Moro 2, I-00185 Roma, Italy\\
$^2$ Departamento de Teoria de la Materia Condensada - 
Instituto de Ciencia de Materiales de Madrid (CSIC)
Cantoblanco, E-28049 Madrid, Spain}
\maketitle

\begin{abstract}
We propose an interpretation for the hump observed in the optical 
conductivity at 
or below a few hundreds of cm$^{-1}$, in overdoped cuprates like the 
electron-doped ${\rm Nd_{2-x}Ce_xCuO_{4-y}}$ at $x \gtrsim 0.15$ and the 
hole-doped ${\rm Bi_2Sr_2CuO_6}$ and ${\rm La_{2-x}Sr_xCuO_4}$.
This interpretation is based on the 
direct excitation of charge collective modes, which become nearly critical in 
the proximity to a charge-ordering instability. The nearly critical character 
of these excitations entails a peculiar temperature dependence and a 
pseudo-scaling form of the lineshapes, which are in agreement with the 
experimental data.
\end{abstract}


\begin{multicols}{2}
The extension of optical conductivity experiments to very low frequencies in 
the range of the very far infrared ($\omega \gtrsim 10$ cm$^{-1}$) has 
uncovered unexpected intriguing features. In particular recent measurements of 
$\sigma(\omega)$ on a slightly overdoped ${\rm Bi_2Sr_2CuO_6}$ 
(Bi2201) sample 
\cite{LUPI2} revealed a strong absorption peak at a few hundreds of cm$^{-1}$ 
narrowing and softening upon decreasing the temperature and well 
separated from a much weaker and narrower Drude term. The extrapolation to 
zero frequency of this latter spectral contribution matches well with the d.c.
value $\sigma(\omega=0,T)$ as obtained from transport data. Althought this is 
the only case where a separate narrow Drude term is directly observed, there 
are other cases both in hole- and electron-doped cuprates, like
${\rm La_{2-x}Sr_xCuO_4}$ (LSCO, ${\rm x=0.184,\, 0.22}$)  \cite{STARTSEVA} 
and ${\rm Nd_{2-x}Ce_xCuO_{4-y}}$ at $x \gtrsim 0.15$ (NCCO) 
\cite{LUPI1,SINGLEY}, where a strongly temperature dependent peak at a finite 
frequency $\Omega_{Max}$ is observed. Moreover, in LSCO, for 
$\omega<\Omega_{Max}$, 
$\sigma(\omega)$ decreases and moves away from the substantially higher
d.c. value. This indicates that a Drude term, which is beyond the lowest 
measured frequency, must intervene to restore a high value of $\sigma$ 
consistently with the zero-frequency data and with the marked metallic 
character of these overdoped systems \cite{nota1}. 

We find that the (direct or indirect) observation of these two separated 
spectral features (the peak at $\Omega_{Max}\lesssim 100-200$ cm$^{-1}$ and 
the narrow Drude term) calls both for an extensive experimental investigation
at very low frequencies and a careful reanalysis of the theoretical 
framework. In this latter regard, the presence of two well separated 
features, with different temperature dependence (at least in Bi2201)
shows that a single-fluid picture, with a strongly frequency-dependent 
inverse scattering time $1/\tau(\omega)$ is inadequate. In particular the 
above experiments show that (at least in some cases) the broad, slowly 
decaying ($\sim 1/\omega$) optical absorption cannot be attributed to 
strongly interacting carriers responsible for
an {\it anomalous} Drude peak at zero 
frequency, and rather arises from finite-frequency excitations clearly 
distinct from much less interacting (and not anomalous) Drude carriers. The 
interpretation of optical spectra at somewhat higher frequency 
$\omega>200-300$ cm$^{-1}$, in terms of quasiparticles (QP's) scattered by 
(spin) collective modes (CM's) \cite{CARBOTTE,CHUBUKOV}, seems hardly viable 
to account for the peculiar behavior of $\sigma(\omega)$ in the above systems 
at $\omega\lesssim 100-200 {\rm cm}^{-1}$. In particular, as we checked (see 
below), and according to previous results \cite{PUCHKOV,KAUFMANN}, overdamped 
modes (as the spin waves would be in the markedly metallic optimally and 
over-doped materials) cannot produce the pronounced Drude-dip-hump 
structure of the spectra.

We propose, instead, that the finite-frequency peak arises from the 
absorption due to the excitation of pairs of CM's directly coupled to the 
electromagnetic (e.m.) field. These CM's are i) abundant at low frequency and 
ii) strongly temperature dependent, because they are critical modes of a 
(hidden) criticality ending at $T=0$ into a quantum critical point (QCP) 
associated to a charge-ordering (CO) instability \cite{COQCP} 
at a critical doping $x_c$ close to 
optimal doping \cite{QCP}. We focus on the low-frequency optical response of 
systems like the above NCCO and Bi2201, which are in the optimally or 
slightly over-doped regions, where a simpler description in terms of QP's 
scattered by the (nearly) critical charge fluctuations $\rho_{\bf q}$, 
associated with the nearby critical line $T_{CO}(x)\approx T^*$, is 
reasonable. We introduce the phenomenological Hamiltonian
\begin{equation}
{\cal H}=\sum_{{\bf k},\sigma}\xi_{\bf k}
c^\dagger_{{\bf k},\sigma}c_{{\bf k},\sigma}
+g\sum_{{\bf k},{\bf q},\sigma} c^\dagger_{{\bf k}+{\bf q},\sigma}
c_{{\bf k},\sigma}\rho_{\bf q},
\label{hamilt}
\end{equation}
where $c^{(\dagger)}_{{\bf k},\sigma}$ are the QP operators, $\xi_{\bf k}$ is 
the QP band and $g$ is the QP-CM coupling constant. While the actual form of 
$\xi_{\bf k}$ is of minor importance, the form of the CM (retarded) 
propagator is quite relevant, 
\begin{equation}
D^R({\bf q},\omega)=
\left[\Gamma+\nu|{\bf q}-{\bf q}_c|^2 -{\rm i}\omega\right]^{-1}.
\label{critprop}
\end{equation}
Such a form, which pertains to a Gaussian QCP with dynamical critical 
index $z=2$, was previously derived within a microscopic Hubbard-Holstein 
model with long-range Coulomb forces \cite{COQCP}, but is rather generic, as 
it can be obtained from other microscopic models with, e.g.,
 antiferromagnetic \cite{AFM} or excitonic \cite{BBG} interactions. The 
``mass term'' $\Gamma$ determines the proximity to the CO instability and 
plays a crucial role since its peculiar temperature dependence 
$\Gamma \approx \max \left[ a(x-x_c), bT \right]$ is a signature of 
quantum critical fluctuations \cite{notaxi} and has observable 
consequences in the optical spectra. Taking all lengths in units
of the lattice spacing, the CM ``bandwidth'' is controlled by the 
energy scale $\nu$ (of the order of the QP Fermy energy $E_F$), which is the 
inverse of a characteristic relaxation time. The wave vector ${\bf q}_c$ sets 
the wavelenght of the most critical charge density fluctuations, 
$\lambda_{CO}\sim |{\bf q}_c|^{-1}$. 

\vspace{-0.75truecm}
\begin{figure}
{\psfig{figure=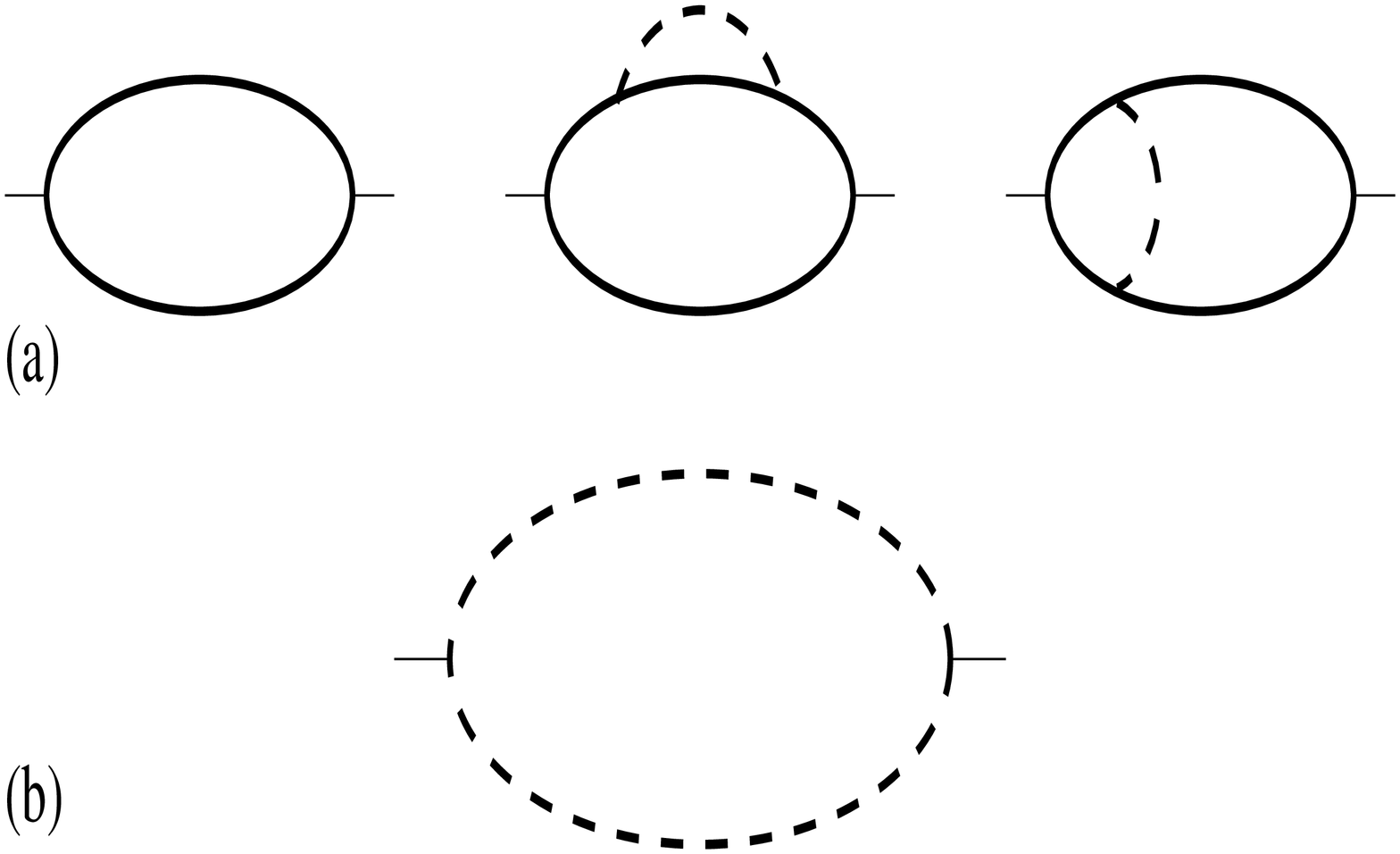,width=5.5cm,angle=-90}}
\end{figure}
\vspace {-1truecm}
{\small FIG. 1. (a) QP (full line) response to the e.m. field, dressed 
by nearly critical CM's (dashed line) at the lowest order in perturbation
theory. (b) CM (dashed line) response to the e.m. field, resulting from
the phenomenological coupling (\ref{gammarho}).}
\vspace{0.3truecm}

Within the model (\ref{hamilt},\ref{critprop}) we calculate the optical 
conductivity arising from (i) the response of the fermionic QP's which are 
scattered by the (nearly) critical CM's [Fig. 1(a)] and (ii) the response of 
the CM's themselves [Fig. 1(b)], which couple to the e.m. vector potential 
via an effective phenomenological interaction 
\begin{equation}
{\cal H}_{A\rho}=\bar g\sum_{{\bf Q},{\bf q}}
\rho_{{\bf q}+{\bf Q}\over 2}\rho_{-{\bf q}+{\bf Q}\over 2} 
{\bf A_Q}\cdot {\bf w}({\bf q},{\bf Q}), 
\label{gammarho}
\end{equation} 
where ${\bf w}({\bf q},{\bf Q})$ is a vectorial matrix element. The 
coupling term (\ref{gammarho}) 
deserves some discussion. We are primarily interested in the effects of 
{\it nearly critical} charge fluctuations, which in the CO scenario occur at 
finite wave vectors ${\bf q} \approx {\bf q}_c$. Therefore the e.m. 
excitation of a single CM at $\vert {\bf q }\vert \approx 0$ is of little 
interest at low energy and we discard it. On the other hand, the simultaneous 
excitation of two CM's with momenta ${\bf q}\approx \pm {\bf q}_c$ 
[Fig. 1(b)] becomes relevant. In this case the actual expression of 
${\bf w}({\bf q},{\bf Q})$ is immaterial since for  
${\bf q}\approx \pm {\bf q}_c$ this vertex only provides a numerical 
coefficient, which can be reabsorbed in $\bar g$.
Of course, these processes can only occur in 
the presence of an effective coupling between the transverse e.m. field and 
the longitudinal charge CM's, which, however, is not problematic. First of 
all, the charge modulation superimposed to the uniform ionic background can 
create a dipolar moment coupling to the e.m. field \cite{EK}. Secondly 
extrinsic 
disorder (from impurities) and/or intrinsic disorder (possibly from a nearly 
static short-range CO texture) can render the very concept of longitudinal 
and transverse excitations meaningless. Finally, and most importantly, the 
direct e.m. excitation of pairs of longitudinal optical phonons via 
quadrupolar coupling or via anharmonic effects is a quite well-known 
phenomenon, which can even acquire a dominant spectral weight close to 
(structural) phase transitions \cite{LIBRO}. 

The QP response, including the standard self-energy and vertex corrections 
due to the CM's, and the direct excitation of the CM's, act as separate 
channels in parallel, and are represented by the diagrams of Fig. 1(a) and 
(b) respectively. For $\omega\ll E_F$, the QP and CM lowest-order corrections
to the current-current correlation functions have the common form
\begin{equation}
\chi_{QP,CM}=\frac{2C_{QP,CM}}{\pi \bar{\Omega}}
\left\lbrack\phi'_{QP,CM}({\omega \over \bar{\Omega}})+ {\rm i}
\phi''_{QP,CM}({\omega \over \bar{\Omega}}) \right\rbrack,
\label{chiqpcm}
\end{equation}
where $C_{QP,CM}$ are constants, and $\bar{\Omega}\sim\Gamma$. 
The imaginary parts 
$\phi''$ are calculated evaluating the corresponding expressions
of the last two diagrams of Fig. 1(a)
at $T=0$, while maintaining the relevant temperature dependence of 
$\bar \Omega\sim \Gamma$. We numerically checked that this approximation 
induces minor quantitative corrections at temperatures and frequencies of 
interest here, while allowing for the explict derivation of the final 
expressions. The real parts $\phi'$ are obtained by a Kramers-Kr\"onig 
transformation, yielding, for $z>0$,
\begin{eqnarray}
\phi''_{QP}(z)&=&-\left\lbrack \frac{1}{z^2}\arctan z -\frac{1}{z}
+\frac{1}{2z}\ln (1+z^2)
\right\rbrack, \label{phiqp2}\\
\phi'_{QP}(z)&=&-\left\lbrack \frac{1}{2z} \arctan z 
+\frac{1}{4z^2} \ln (1+z^2)
\right\rbrack; \label{phiqp1}\\
\phi''_{CM}(z)&=&\frac{1}{z^2+4}\left\lbrack \arctan z -\frac{z^2+2}{2z}
\ln (1+z^2) \right\rbrack,  \label{phicm2}\\
\phi'_{CM}(z)&=&-\frac{1}{z^2+4}\left\lbrack \frac{z^2+2}{z} \arctan z + 
\frac{1}{2} \ln (1+z^2)
\right\rbrack.  \label{phicm1}
\end{eqnarray}
The contribution of the CM's to the complex conductivity is obtained by
directly inserting the corresponding current-current response function
from the diagram in Fig. 1(b) into the Kubo formula yielding 
\begin{equation}
\sigma_{CM}(\omega)=\frac{{\rm i}\chi_{CM}(\omega)}{\omega}.
\label{sigmacm}
\end{equation}
On the other hand, it is well known \cite{GOETZE} that 
$\chi_{QP}$, which is the lowest-order correction to the optical response of 
a perfect metal, provides a valid expression for $\sigma(\omega)$
in the high-frequency regime only ($\omega\tau\gg 1$). This difficulty
is customarily overcome by expressing the 
the QP contribution to the conductivity in terms of
a complex memory function $M$ \cite{GOETZE}
\begin{equation}
\sigma_{QP}(\omega)=\frac{\omega_p^2}{4\pi}\, 
\frac{1}{M_{QP}(\omega,T)-{\rm i}\omega}.
\label{sigmaqp}
\end{equation}
A perturbative approach to $M$ is now viable, even at low frequencies
leading to $M_{QP} \approx {\rm i}\omega 4\pi\chi_{QP}/\omega_p^2$. 
For $\sigma_{QP}$ this corresponds to 
a RPA resummation of the diagrams in Fig. 1(a).
Here $\omega_p$ is the plasma frequency, while ${\rm Re} M$ and 
$-{\rm Im} M/\omega$ play the role of the inverse scattering time, and ``mass 
renormalization'' (which reduces the low-frequency absorption), respectively.

The total absorption arises from 
the parallel of the two separate contributions given in Eqs. 
(\ref{sigmacm}),(\ref{sigmaqp}) $\sigma=\sigma_{QP}+\sigma_{CM}$. A similar 
two-component conductivity was discussed within a one-dimensional 
Kondo-lattice model \cite{EK}, where, however, no critical behavior was 
present. Within this scheme, the scattering of the QP's with impurities, with 
a scattering time $\tau_{imp}$, is in series with the scattering 
of the QP's with the CM's and, according to the Matthiessen's rule,
is accounted for by taking $M_{QP}\to M_{QP}+\tau_{imp}^{-1}$.


\begin{figure}
{\psfig{figure=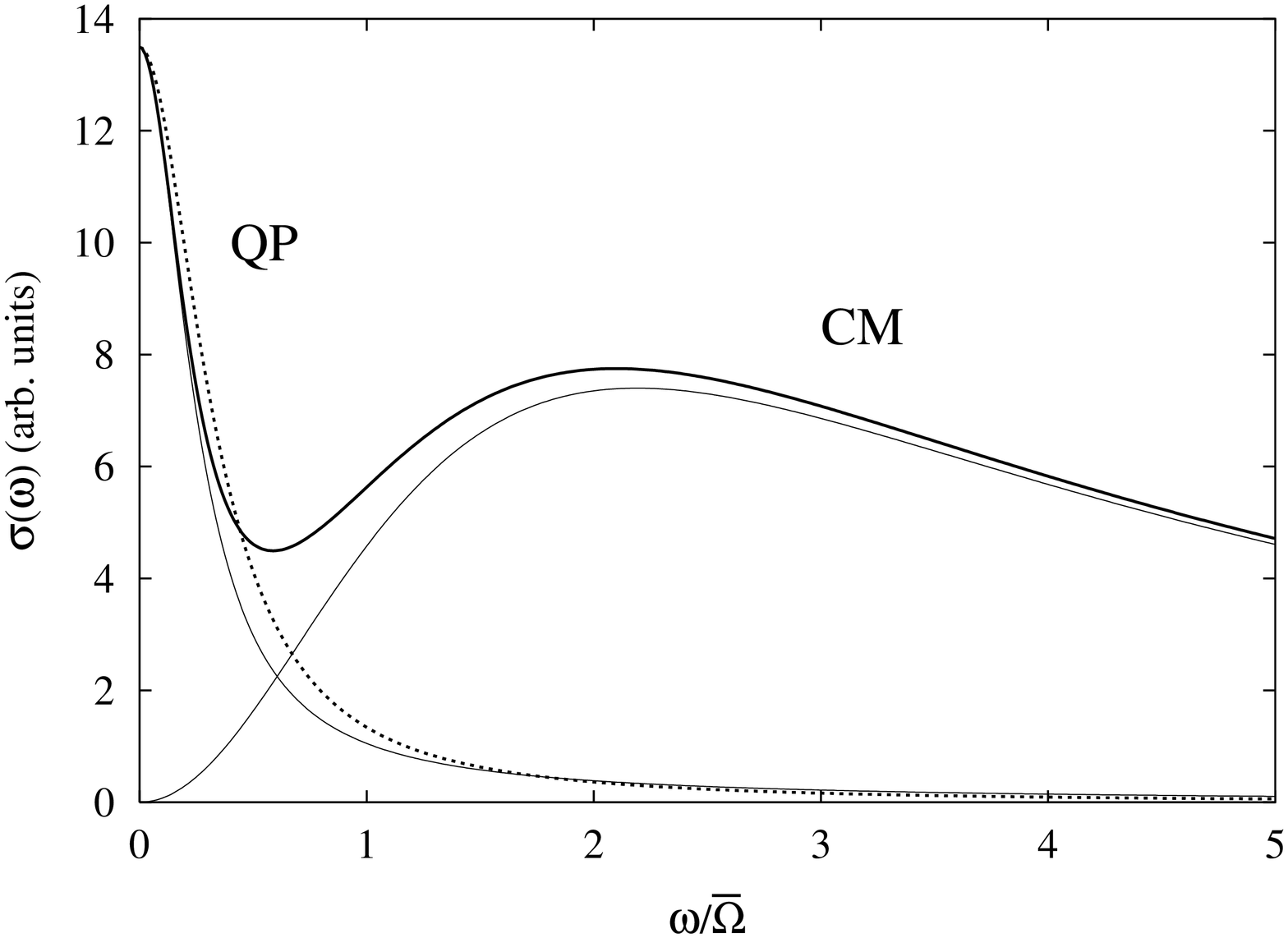,width=7.5cm,angle=-90}}
\end{figure}
\vspace {-0.5truecm}
{\small FIG. 2. Thick line: typical optical absorption 
from collective-modes [CM, eq. (\ref{sigmacm})] and
quasiparticles [QP, eq. (\ref{sigmaqp})], including impurity
scattering with $\tau_{imp}^{-1}=\bar{\Omega}/3$. The
separate contributions are shown as thin lines. An ordinary Drude
peak with the same $\tau_{imp}^{-1}$ is shown for comparison (dotted
line). Notice the depletion of spectral weight at $\omega \lesssim 
\bar{\Omega}$ due to the QP-CM scattering.}
\vspace{.3cm}

A typical optical spectrum resulting from our calculations is
displayed in Fig. 2. In the absence of disorder, the QP contribution 
consists of a (nearly) delta-like absorption at zero frequency 
[essentially arising from the first diagram of Fig. 1(a)] plus a long tail
 which  monotonically  decreases with $\omega$ according
to Eqs. (\ref{phiqp2}) and (\ref{phiqp1}).
In agreement with previous analyses \cite{PUCHKOV,KAUFMANN}
this second contribution has a weight proportional to the  QP-CM coupling
$g^2$, while its width is controlled by $\bar{\Omega}$. Impurity
scattering gives the zero-frequency peak a finite width, which,
mostly due to mass-renormalization effects (the ImM contribution), is
{\it less} than the expected 
Drude value $\tau_{imp}^{-1}$ (see dotted line in
Fig. 2), indicating that
the QP-CM scattering transfers some spectral weight 
from the region $\omega \lesssim \tau_{imp}^{-1}$ to the region $\omega
\sim \bar{\Omega}$. Roughly speaking this occurs because the QP's
dressed with the CM's acquire spectral weight around the typical CM frequencies
(and in some sense also a partial CM character) thereby becoming less sensitive
to the single-particle scattering due to impurities. As a result,
it turns out that the coupling with the CM's narrows the Drude peak
for $\omega \lesssim \tau_{imp}^{-1}$
and the system appears somewhat cleaner than it would be in the absence
of QP-CM coupling. However,
even taking a very large $g^2$, we fail in producing a non-monotonic 
behavior with maxima at finite frequency. 
Moreover, we see that for all reasonable choices of the parameters,
the broad tail is very weak compared to the large finite-frequency
absorption seen in the experiments.
By no means one can reconcile the low and narrow Drude peak with
the higher and broader hump within the QP component only.
The contribution from the diagram in Fig. 1(b) is therefore crucial
to reproduce the experimental hump and supports the introduction of the
phenomenological interaction (\ref{gammarho}). Indeed
the convolution of two overdamped 
CM's having the critical form of Eq. (\ref{critprop}), leads to an 
absorption with a (broad) maximum at 
$\omega = \Omega_{Max}=2\bar\Omega\sim 2\Gamma$.  
Theoretically, this quantity should scale proportionally 
to $T$ at high temperature, in the quantum critical regime, and saturate to 
a constant value $\propto (x-x_c)$ in the low-temperature quantum 
disordered regime, as ruled by the proximity to a QCP. The critical character 
of the propagator (\ref{critprop}) entails a scaling form in the 
frequency dependence of 
$\chi(\omega)={\rm const}\times\phi (\omega/\Omega_{Max})/\Omega_{Max}$. As a 
consequence, the corresponding finite-frequency features in $\sigma(\omega)$ 
display a scaling behavior, when plotted as a 
function of $\omega/\Omega_{Max}$ (see Fig. 3) \cite{notascaling}.

\begin{figure}
{\psfig{figure=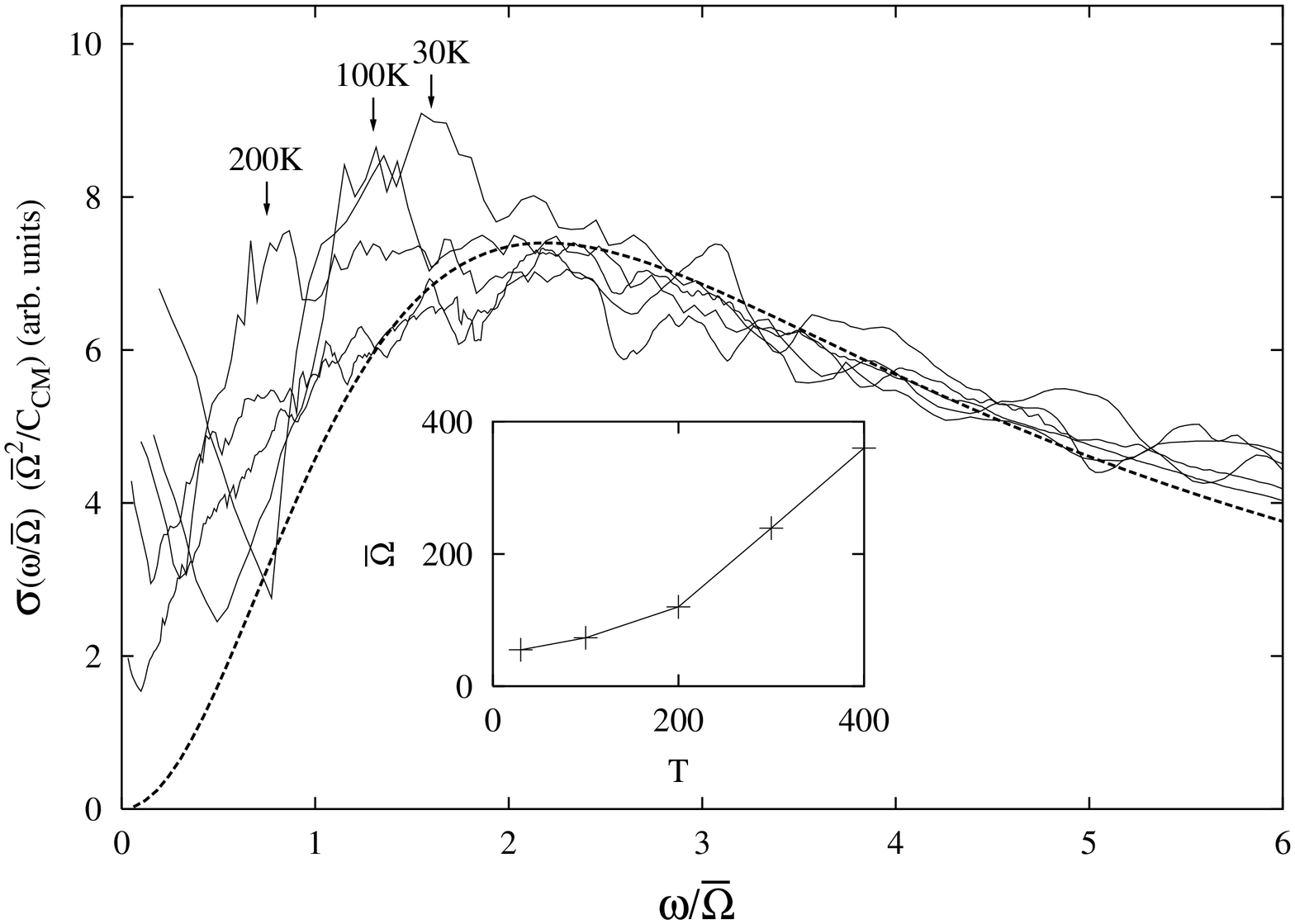,width=8.5cm,angle=-90}}
\end{figure}
\vspace {-0.5truecm}
{\small FIG. 3. Scaled optical absorption
$\sigma(\omega/\bar{\Omega})(\bar{\Omega}/\bar{\omega}_p)^2$ for slightly 
overdoped Bi2201 at different temperatures. The absorption peaks all fall 
on a single theoretical curve (dashed), given by Eq. (6). The deviations 
from the scaling curve in the low frequency part come from the Drude
contribution  (data courtesy of S. Lupi {\em et al.}). The inset shows the
behavior of the fit parameter $\bar{\Omega}$ vs $T$, characteristic of 
the crossover from a quantum critical to a quantum disordered regime 
(see text).}
\vspace{.3cm}

The peculiar $\omega$ and $T$ dependence of $\sigma$ described above is 
indeed seen in the optical spectra of the electron-doped NCCO at 
$x \gtrsim 0.15$ \cite{LUPI1,SINGLEY} and of the hole-doped Bi2201 
\cite{LUPI2} in the non-superconducting phase. 
Despite the limited amount of available data, we find quite 
encouraging that our approach is in remarkable agreement with experiments 
whenever a comparison is possible (i.e. in experiments carried out down to 
very low frequencies $\omega\gtrsim 10 \,{\rm cm}^{-1}$). 
Specifically, we use our theory to fit the spectra 
of Bi2201, reported in Ref. \cite{LUPI2}. According to the above
discussion, we analyze the Drude part and 
the  anomalous hump separately. The latter can be fitted
 adjusting the strength of the (e.m. field)-CM coupling 
$\bar g$ [more exactly the constant $C_{CM}$ in 
Eq. (\ref{chiqpcm})] determining the weight of the absorption from CM's
and the CM characteristic frequency scale $\bar\Omega\sim\Gamma$. 
As it can be seen from Fig. 3, where
 we plot the experimental and theoretical curves in the scaled 
variable $\omega/\bar\Omega$ the scaling collapse of the
curves into a single curve and the fit of the spectra at different
temperatures is satisfactory.  A similar agreement is found in the case 
of the NCCO and LSCO data of Refs. \cite{SINGLEY,STARTSEVA} respectively.

A direct inspection of the non-rescaled data in Fig. 3 of Ref.
\cite{LUPI2} shows the presence around 100 cm$^{-1}$ of a narrow feature 
(likely of phononic origin) which narrows but does not shift upon
lowering the temperature. This extrinsic feature  (marked by
vertical arrows in Fig. 3) obviously does
not possess any scaling character, which   
explains  the deviations from the universal curve 
in the low-frequency region. 

It is quite remarkable, that $\bar \Omega=\Omega_{Max}/2$ 
as obtained from the fits and reported in the inset of Fig. 3, 
is linear in temperature and then saturates at a finite value at
low temperatures as expected for $\Gamma$ from quantum criticality. A similar 
quantum-critical dependence has recently been observed by ARPES in the 
scattering rate of QP's in optimally-doped Bi2212 \cite{VALLA}, indicating 
the still not understood but remarkably general character of this behavior. 

We now turn our attention to the small and narrow 
zero-frequency peak. Setting the CM-QP coupling to zero,
as shown in Ref. \cite{LUPI2}, this  peak
could be easily fitted as an ordinary Drude peak with two 
parameters, the plasma
frequency $\omega_p$, which provides the intensity, and the 
scattering rate $\tau_{imp}^{-1}\approx 30-50 cm^{-1}$, 
which controls the width. These values for $\tau_{imp}^{-1}$
are fairly small for such chemically complex compounds. However, 
according to the discussion of Fig. 2, the introduction of
a moderate QP-CM coupling tends to reduce the width of the Drude peak
thereby rendering the observed Drude width compatible with
$\tau_{imp}^{-1}$'s as large as $100-200 cm^{-1}$.

The comparison between theory and experiments is only meaningful at low 
frequency, since our theory is based on a general analysis of quantum 
critical phenomena, which fails in describing the non-critical high-frequency 
(mid-infrared, charge-transfer,...) parts of the spectra. Nevertheless, the 
(approximate) scaling form and temperature dependence of $\sigma(\omega)$ are 
non-trivial robust features of our results. We notice that the pseudoscaling 
properties of the softening of the finite-frequency feature upon reducing the 
temperature is hardly interpreted within other (more traditional) mechanisms, 
like the binding due to impurity states \cite{SINGLEY}, and indicates that 
$T$ {\em is the only relevant energy scale}, determining both the position 
and the shape of the ``hump'' spectral feature, in a broad temperature range. 
At the 
moment, we are not aware of any well-established alternative explanation, 
other than the direct absorption from nearly critical CM's.
This is particularly important, not only because it provides an 
explanation for the spectral features of specific compounds, but also, and 
most importantly, because it supports the peculiar role of critical charge 
fluctuations in ruling the physical properties of the cuprates.


{\em Acknowledgments.} We acknowledge stimulating discussions
with P. Calvani, M. Capizzi, C. Castellani, and S. Lupi, whom we also
thank for providing us the data of the optical conductivity in Bi2201.


\end{multicols}

\begin{references}
\bibitem{notaxi} This behavior is an approximate way to interpolate between
	the two asymptotic behaviors $\Gamma \approx a(x-x_c)$ or 
        $\Gamma \approx bT$ (with $a$ and $b$ non-universal constants) 
        characterizing the quantum disordered and the quantum critical 
        regimes respectively.
\end{references}
\end{document}